\documentclass[conference,dvipsnames]{IEEEtran}
\usepackage{geometry}
\usepackage[T1]{fontenc}
\usepackage[utf8]{inputenc}
\usepackage{mathtools,amssymb,amsthm}
\usepackage{csquotes}
\usepackage{nicefrac}
\usepackage{subcaption}
\usepackage{booktabs}
\usepackage{enumitem}
\usepackage{siunitx}
\usepackage{url}
\usepackage[noadjust]{cite}
\usepackage{xcolor}
\usepackage{tikz}
\usepackage{pgfplots}
\usepackage[bookmarks=false]{hyperref}

\pgfplotsset{compat=1.14}
\usetikzlibrary{arrows.meta, backgrounds, calc, matrix, positioning}

\newcommand{\ket}[1]{\ensuremath{\left|#1\right\rangle}}

\newtheorem{example}{Example}

\begin{document}
\bstctlcite{IEEEnodash:BSTcontrol}
\hypersetup{
	pdftitle={Approximation of Quantum States Using Decision Diagrams},
	pdfsubject={25th Asia and South Pacific Design Automation Conference (ASP-DAC) 2020},
	pdfauthor={Alwin Zulehner, Stefan Hillmich, Igor L. Markov, and Robert Wille},
}
\title{Approximating the Quantum State with Decision Diagrams}
\title{Trading-Off Fidelity and Compactness:\\Approximation of Quantum States Using Decision Diagrams}
\title{\LARGE\textbf{Approximation of Quantum States Using Decision Diagrams}}
\author{Alwin Zulehner$^1$, Stefan Hillmich$^1$, Igor L. Markov$^2$, and Robert Wille$^1$ \\ 
	{\normalsize $^1$Institute for Integrated Circuits, Johannes Kepler University Linz, Austria}\\
	{\normalsize $^2$Department of EECS, University of Michigan, USA}\\
	{\normalsize alwin.zulehner@jku.at, stefan.hillmich@jku.at, imarkov@eecs.umich.edu, robert.wille@jku.at}\\
	{\small \url{http://iic.jku.at/eda/research/quantum/}}}
\date{}

\maketitle

{\small\textbf{Abstract---\ignorespaces
	The computational power of quantum computers poses major challenges to new design tools since representing pure quantum states typically requires exponentially large memory.
	As shown previously, decision diagrams can reduce these memory requirements by exploiting redundancies.
	In this work, we demonstrate further reductions by allowing for small inaccuracies in the quantum state representation.
	Such inaccuracies are legitimate since quantum computers themselves experience gate and measurement errors and since quantum algorithms are somewhat resistant to errors (even without error correction).
	We develop four dedicated schemes that exploit these observations and effectively approximate quantum states represented by decision diagrams.
	We empirically show that the proposed schemes reduce the size of decision diagrams by up to several orders of magnitude while controlling the fidelity of approximate quantum state representations.
}

\section{Introduction}

Quantum computing~\cite{NC:2000} enables exponential speed-ups compared to conventional computers by exploiting quantum mechanical effects such as \emph{superposition}, where qubits can assume a linear combination of their basis states \ket{0} and \ket{1}, and \emph{entanglement}, where qubits may be correlated. 
As a result, the pure state of a quantum system composed of $n$ qubits may represent a superposition of $2^n$~basis~states and corresponding complex amplitudes---promising a greater amount of information per qubit and greater computational power. 
First algorithms following this powerful computing paradigm have been developed in the last century such as Shor's algorithm for integer factorization~\cite{Sho:94} and Grover's algorithm for searching in unsorted databases~\cite{DBLP:conf/stoc/Grover96}. Recently, the development of new quantum algorithms targeting various problems in physical simulation, chemistry, finance, and machine learning~\cite{montanaro2016quantum,preskill2018quantum,coles2018quantum} has gained momentum since significant progress in the physical realization of quantum computers has been achieved by players such as IBM, Google, Microsoft, Intel, Rigetti~\cite{gomes2018quantumcomputing}.  

The development of quantum algorithms and their successful implementation on quantum computers relies on several tasks (such as simulation, state preparation, and verification) that have to be completed on conventional computers. Here, the enormous computational power of quantum computers becomes a liability since the amplitudes describing the \emph{wave function} of the quantum system are often represented by means of a \mbox{$2^n$-dimensional} state vector---implying exponential memory requirements.

In the past, several data structures have been developed to overcome this exponential memory complexity of representing quantum states---including \emph{Matrix Product States} (MPS~\cite{orus2014practical}), \emph{Projected Entangled Pair States} (PEPS~\cite{orus2014practical}), and even neural networks~\cite{DBLP:journals/qip/SchuldSP14}. 
Besides that, \emph{Decision Diagrams}~(DDs~\cite{viamontes2003improving,DBLP:journals/tcad/NiemannWMTD16,AP:06Neu,WLTK:2008}) have been intensely investigated and, in the meantime, have shown great potential for the design tasks outlined above~\cite{AP:06Neu,WLTK:2008,VMH:2007,DBLP:conf/rc/NiemannWD14,niemann2018cliffordt,DBLP:conf/ismvl/WilleGMD09,DBLP:journals/tcad/ZulehnerW18OnePass}---especially for quantum-circuit simulation~\cite{DBLP:books/daglib/0027785,DBLP:journals/tcad/ZulehnerW18,zulehner2019matrix}. This is in line with the successes of conventional decision diagrams such as BDDs~\cite{Bry:86}, KFDDs~\cite{DBLP:journals/tcad/DrechslerB98}, etc.~which got established in numerous design tasks for conventional circuits.

Even with specialized data structures, the memory requirements for the respective representations remain a significant challenge.  
Although decision diagrams allow for a rather compact representation in many cases, we note that approximations of quantum states have not yet been used to reduce memory usage. 
To this end, quantum algorithms are resistant to errors to some extent (even without error correction) and quantum computers experience gate as well as measurement errors.
Hence, allowing for small inaccuracies in the state representation does not (or hardly) affect the result of these probabilistic measurements.
Moreover, current quantum computers are subject to noise~\cite{preskill2018quantum,gambetta2017building} and, thus, do not represent desired quantum states exactly anyway.
These considerations
suggest possibilities for memory compression which has not been investigated thus far.

In this work, we investigate the potential of approximation to obtain more compact DD-based representations of quantum states. 
The key idea is to eliminate nodes from the decision diagram that represent basis states with minuscule contributions to the overall state vector---setting  
small amplitudes to zero.
More precisely, we present several dedicated approximation schemes with linear time complexity in the size of the decision diagram, 
which can reduce its size significantly. 
Tuning hyper-parameters of these schemes ensures a certain fidelity when determining approximations.
Empirical evaluations confirm this behavior and show that the size of the decision diagrams can be reduced by several orders of magnitude while also controlling the fidelity of the resulting approximated state representations.
Given our results, these schemes can be automatically invoked during tasks such as simulation to reduce the size of the decision diagram when it grows too large, similar to garbage collection some programming languages.

The remainder of this paper is structured as follows. 
Section~\ref{sec:background} reviews the background  on quantum states and decision diagrams. 
Afterwards, Section~\ref{sec:gen_idea} describes the main idea of state approximation, whereas Section~\ref{sec:approx_schemes} discusses four dedicated schemes to perform the approximations.
Section~\ref{sec:evaluation} summarizes the evaluation of the presented schemes while Section~\ref{sec:conclusion} concludes the paper.

\section{Background}
\label{sec:background}

This section provides the background to keep the exposition \mbox{self-contained}. It reviews quantum states as well as decision diagrams that can represent them in conventional software.

\subsection{Quantum States}
\label{subsec:quantum-basics}

In the realm of quantum computing, conventional bits are generalized to \emph{quantum bits} (i.e.,~\emph{qubits}). 
While the former can only be in exactly one of the states \(0\) and \(1\), qubits may  assume two  basis states (denoted~\(\ket{0}\) and \(\ket{1}\)) as well, but also any linear combination of them. This is described by \( \ket{\psi} = \alpha_0\cdot\ket{0} + \alpha_1\cdot\ket{1} \) with \emph{amplitudes} \(\alpha_0, \alpha_1 \in \mathbb{C} \) and the normalization constraint \( |\alpha_0|^2 + |\alpha_1|^2 = 1 \).
Qubits with \(\alpha_0\) and \(\alpha_1\) unequal to zero are referred to as being in \emph{superposition}, i.e.,~they are in \enquote{both states} at the same time---one of the main characteristics of quantum computing that enables substantial speed-ups in certain applications due to the resulting potential for considering multiple states at the same time\footnote{Another important characteristic is \emph{entanglement}, where the state of a single qubit may influence the state of another qubit.}.

Unfortunately, the respective amplitudes (\(\alpha_0\) and \(\alpha_1\)) cannot be directly observed on a quantum computer. Instead, they dictate the probability of certain outcomes of a measurement with respect to the corresponding basis states. More precisely, measuring a single qubit in state \mbox{\( \ket{\psi} = \alpha_0\cdot\ket{0} + \alpha_1\cdot\ket{1} \)} yields the output \(\ket{0}\) with probability~\( |\alpha_0|^2 \) or the output~\(\ket{1}\) with probability~\( |\alpha_1|^2 \). After the measurement, the qubit will lose any superposition, i.e., it collapses into a basis state.
For quantum systems composed of more than one qubit, the description is accordingly extended. For example, a system with two qubits has four basis states, i.e., \mbox{\( \ket{\psi} = \alpha_{00}\cdot\ket{00} + \alpha_{01}\cdot\ket{01} + \alpha_{10}\cdot\ket{10} + \alpha_{11}\cdot\ket{11} \)} with the normalization constraint \mbox{\( |\alpha_{00}|^2 + |\alpha_{01}|^2 + |\alpha_{10}|^2 + |\alpha_{11}|^2 = 1 \)}.
Commonly, the description of a quantum state is shortened to a vector containing the amplitudes \( \ket{\psi} = \left[\alpha_{00}, \alpha_{01}, \alpha_{10}, \alpha_{11}\right]^\mathrm{T} \).

\begin{example}
	\label{ex:qubits}
	Consider an arbitrary quantum system with two qubits, which are entangled and in superposition. 
	The state is given as \mbox{\(\ket{\psi} = \nicefrac{1}{\sqrt{2}}\cdot\ket{00} + 0\cdot\ket{01} + 0\cdot\ket{10} +\nicefrac{1}{\sqrt{2}}\cdot\ket{11}\)}.
	This state is valid as the normalization constraint \({\left(\nicefrac{1}{\sqrt{2}}\right)^2 + 0^2 + 0^2 + \left(\nicefrac{1}{\sqrt{2}}\right)^2 = 1}\) is satisfied. 
	The quantum state is represented as a vector \(\psi = \left[ \nicefrac{1}{\sqrt{2}}, 0, 0, \nicefrac{1}{\sqrt{2}} \right]^\mathrm{T}\).
	
	Performing a measurement on this system yields one of the two basis states \(\ket{00}\) or \(\ket{11}\), each with a probability of \mbox{\(\left| \nicefrac{1}{\sqrt{2}}\right|^2 = \nicefrac{1}{2}\)}. 
	After the measurement, the quantum state collapses, i.e.,~the state vector is either \mbox{\(\psi = \left[ 1, 0, 0, 0 \right]^\mathrm{T}\)} or \mbox{\(\psi = \left[ 0, 0, 0, 1 \right]^\mathrm{T}\)}, depending on the measured state.
\end{example}

\subsection{Decision Diagrams}
\label{subsec:bg-dds}

Decision diagrams have been successfully utilized to drastically reduce the memory requirements for representing state vectors in quantum computing~\cite{AP:06Neu,viamontes2003improving,WLTK:2008,DBLP:journals/tcad/NiemannWMTD16}. For example, simulation approaches based on decision diagrams have recently moved into the spotlight since they significantly outperform array-based simulators in cases where redundancies can be exploited (in extreme cases leading to an improvement in runtime from 30 days to 2 minutes~\cite{DBLP:journals/tcad/ZulehnerW18}).

In the context of quantum state representation, decision diagrams identify redundancies in the state vector and provide compaction by sharing structures.
The vector is split into two equal-sized sub-vectors.
This process is repeated until the sub-vector contains a single element only, i.e.,~one split for every qubit.
If identical \mbox{sub-vectors} occur in the process, this redundancy is exploited by re-using (sharing) the same structure in the resulting decision diagram.

More precisely, consider an arbitrary quantum system with \(n\) qubits which are labeled \(q_0 , q_1 , \ldots, q_{n-1} \), where \( q_0 \) represents the most significant qubit of the quantum state.
The state of \(q_0\) splits the state vector in half: the first \(2^{n-1}\) entries correspond to \(q_0=\ket{0}\), while the other entries correspond to \(q_0=\ket{1}\).
The decision diagram structure represents this decomposition through the node labeled \(q_0\) with two successor nodes that represent the aforementioned sub-vectors.
Commonly, the left node denotes the 0-successor where as the right node denotes the 1-successor.
This decomposition is recursively applied until the remaining sub-vectors have size \(1\), i.e.,~they represent a single complex number.
In the process, two equivalent \mbox{sub-vectors} are represented by the same node, which reduces the memory requirements by exploiting redundancies.
Nodes representing \(q_{n-1}\) have a single terminal for their left and right successors.
This is achieved by storing common factors in the edge weights.
To determine the value of an amplitude, the edge weights of the path representing said amplitude are multiplied.

	\begin{figure}[tbp]
		\centering
		\begin{subfigure}[t]{0.4\linewidth}
			\centering
			\begin{tikzpicture}
				\matrix[matrix of math nodes, left delimiter={[},right delimiter={]}, inner xsep=0] (vector) {
					0\\
					\nicefrac{2}{\sqrt{10}}\\
					0\\
					\nicefrac{2}{\sqrt{10}}\\
					\nicefrac{1}{\sqrt{10}}\\
					0\\
					0\\
					{\nicefrac{-1}{\sqrt{10}}}\\
				};
				
				\begin{scope}[on background layer]	
					\node[left=0.75cm of vector-1-1.center] {\(\ket{000}\)};
					\node[left=0.75cm of vector-2-1.center] {\(\ket{001}\)};
					\node[left=0.75cm of vector-3-1.center] {\(\ket{010}\)};
					\node[left=0.75cm of vector-4-1.center] {\(\ket{011}\)};
					\node[left=0.75cm of vector-5-1.center] {\(\ket{100}\)};
					\node[left=0.75cm of vector-6-1.center] {\(\ket{101}\)};
					\node[left=0.75cm of vector-7-1.center] {\(\ket{110}\)};
					\node[left=0.75cm of vector-8-1.center] {\(\ket{111}\)};
				\end{scope}
				\end{tikzpicture}
			\caption{Vector representation}
			\label{fig:statevectorvector}
		\end{subfigure}
		\qquad
		\begin{subfigure}[t]{0.45\linewidth}
			\centering
			\begin{tikzpicture}[terminal/.style={draw,rectangle,inner sep=0pt}]	
			\matrix[matrix of nodes,ampersand replacement=\&,every node/.style={draw,circle,inner sep=0pt,minimum width=0.5cm,minimum height=0.5cm},column sep={1cm,between origins},row sep={1cm,between origins}] (qmdd2) {
								\& |(m1)|$q_0$ 											\\
				|(m2a)| $q_1$	\&                                  \& |[xshift=-0.5cm](m2b)| $q_1$ 						\\
				|(m3a)| $q_2$	\& |[xshift=0.25cm](m3b)| $q_2$ 					\& |(m3c)| $q_2$ 	\\
								\& |[terminal] (t3)| $1$ 									\\
			};
			
			\draw[thick] ($(m1)+(0,0.7cm)$) -- (m1) node[right, midway]{$\frac{2}{\sqrt{10}}$};
			
			\draw (m1) -- ++(240:0.6cm) -- (m2a);
			\draw[thick] (m1) -- ++(300:0.6cm) node[right, midway] {$\frac{1}{2}$} -- (m2b);
			
			\draw (m2a) -- ++(240:0.6cm) -- (m3a);
			\draw (m2a) -- ++(300:0.6cm) -- (m3a);
			
			\draw (m2b) -- ++(240:0.6cm) -- (m3b);
			\draw[thick] (m2b) -- ++(300:0.6cm) node[right, midway] {$-1$} -- (m3c);
			
			\draw (m3a) -- ++(240:0.4cm) node[below, xshift=0.5pt, inner sep=0,font=\tiny] {$0$};
			\draw (m3a) -- ++(300:0.6cm) -- (t3);
			
			\draw (m3b) -- ++(240:0.6cm) -- (t3);
			\draw (m3b) -- ++(300:0.4cm) node[below, xshift=0.5pt, inner sep=0,font=\tiny] {$0$};
					
			\draw (m3c) -- ++(240:0.4cm) node[below, xshift=0.5pt, inner sep=0,font=\tiny] {$0$};
			\draw[thick] (m3c) -- ++(300:0.6cm) -- (t3);
			\end{tikzpicture}
			\caption{Decision diagram}
			\label{fig:dd-statevector}
		\end{subfigure}
		\caption{Given quantum state}
	\end{figure}
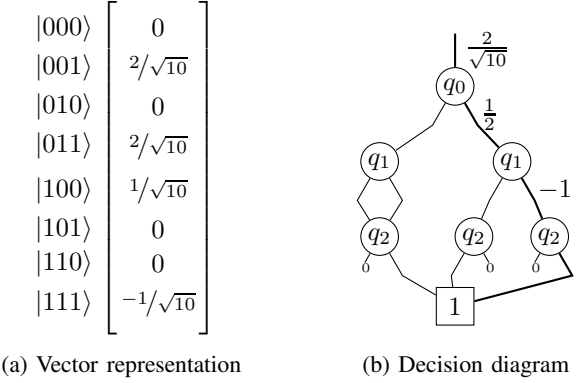
	
\begin{example} 
	Consider the state vector in Fig.~\ref{fig:statevectorvector}. 
	The corresponding basis state to each entry of the vector is written to the left.
	The decision diagram representing the same state as the vector is depicted in Fig.~\ref{fig:dd-statevector}.
	The amplitude of any state is accessed by following the path in the decision diagram and multiplying the corresponding edge weights.
	For the state \ket{111}, the path is bolded (\(q_0 = 1\), \(q_1 = 1\), \(q_2 = 1\)) and the amplitude is \(\nicefrac{2}{\sqrt{10}} \cdot \nicefrac{1}{2} \cdot (-1) \cdot 1 = \nicefrac{-1}{\sqrt{10}} \).
\end{example}

\section{General Idea}\label{sec:gen_idea}
This section describes and illustrates the general idea of the proposed approximation approach. Naturally, designers are interested in an as compact as possible representation of the quantum states generated, e.g.,~by a quantum algorithm---a challenging task given that straightforward representations of states are exponential in size. As reviewed in the previous section and as already utilized in the conventional domain for decades, decision diagrams provide a well suited means for this purpose (by reducing the complexity through sharing of redundancies). In contrast to their conventional counterparts, quantum computations provide another useful property that allows for more compact representations by means of decision diagrams: 
Quantum algorithms are resistant to errors to some degree.
Additionally, quantum computers suffer from gate as well as measurement errors anyway.

At each run, a quantum computer generates a bitstring (representing the measured basis state) which is affected by the aforementioned errors.
More precisely, while the respective amplitude~$\alpha_i$ of a basis state \(\ket{i}\) dictates the probability of the measurement, small changes in the amplitude introduced by some error or approximation scheme do not (or hardly) affect the measurement due to the error resistance of the quantum computation. 
Because of that, perfect precision of a state representation is not necessary in most cases anyway. 
Besides that, quantum computing itself suffers from noise which additionally renders a precise representation unnecessary in many cases.

Exploiting the need of quantum computing to operate in noisy environments with gate and measurement errors offers further potential for more compact decision diagrams: Rather than aim to precisely represent the quantum state with amplitudes~$\alpha_i$ for \emph{all} possible basis states~$\ket{i}$, the quantum state is \emph{approximated}. 
The DD-based representation can be approximated in two ways to become more compact.
First, \emph{similar} nodes can merged to a single node. This is implicitly implemented by considering complex numbers with a difference below a given tolerance value as equal.
Second, basis states which do not contribute significantly to the quantum state (i.e.,~states whose amplitudes are close to zero) are eliminated. Removing these nodes from the decision diagram sets the corresponding amplitudes to zero. 
Since quantum states are represented by unit-norm complex vectors, the remaining amplitudes are scaled (dividing by the magnitude of the approximated vector) such that the normalization constraint $\sum_{i=0}^{2^n-1} |\alpha_i|^2 = 1$ remains satisfied.

\begin{example}
	Assume that basis states $\ket{i}$ with $|\alpha_i|^2 < \nicefrac{1}{4}$ are indistinguishable from noise.
	Then the quantum state provided in Fig.~\ref{fig:statevectorvector} is approximated by setting the amplitudes $\alpha_{100}$ and $\alpha_{111}$ to~\(0\). 
	To normalize the state, we scale the remaining amplitudes by dividing them by \(\sqrt{|\nicefrac{2}{\sqrt{10}}|^2 + |\nicefrac{2}{\sqrt{10}}|^2} = \nicefrac{2}{\sqrt{5}}\)---leading to the state shown in Fig.~\ref{fig:dd-statevector-approximated-vector}.
	
	The decision diagrams representing the original and the approximated quantum state are shown in Fig.~\ref{fig:dd-statevector} and Fig.~\ref{fig:dd-statevector-approximated-dd}, respectively. 	
	The original paths following the right successor of the \(q_0\) node each have a probability of \(\left|\nicefrac{\pm 1}{\sqrt{10}}\right|^2 = \nicefrac{1}{10}\) and, therefore, are removed.
	This reduces the size of the decision diagram from six nodes to three, yielding an approximation that constitutes an acceptable trade off for certain applications, where probabilities below a certain threshold are useless anyway.
\end{example}

\begin{figure}
	\centering
	\begin{subfigure}[b]{0.45\linewidth}
		\centering
		\scalebox{1}{\begin{tikzpicture}
			\matrix[matrix of math nodes, left delimiter={[},right delimiter={]}, inner xsep=0] (vector) {
				0\\
				\nicefrac{1}{\sqrt{2}}\\
				0\\
				\nicefrac{1}{\sqrt{2}}\\
				0\\
				0\\
				0\\
				0\\
			};
			\begin{scope}[on background layer]	
				\node[left=0.75cm of vector-1-1.center] {\(\ket{000}\)};
				\node[left=0.75cm of vector-2-1.center] {\(\ket{001}\)};
				\node[left=0.75cm of vector-3-1.center] {\(\ket{010}\)};
				\node[left=0.75cm of vector-4-1.center] {\(\ket{011}\)};
				\node[left=0.75cm of vector-5-1.center] {\(\ket{100}\)};
				\node[left=0.75cm of vector-6-1.center] {\(\ket{101}\)};
				\node[left=0.75cm of vector-7-1.center] {\(\ket{110}\)};
				\node[left=0.75cm of vector-8-1.center] {\(\ket{111}\)};
			\end{scope}
		\end{tikzpicture}}
		\caption{Vector representation}
		\label{fig:dd-statevector-approximated-vector}
	\end{subfigure}\quad
	\begin{subfigure}[b]{0.45\linewidth}
		\centering
		\scalebox{1}{\begin{tikzpicture}[terminal/.style={draw,rectangle,inner sep=0pt}]	
		\matrix[matrix of nodes,ampersand replacement=\&,every node/.style={draw,circle,inner sep=0pt,minimum width=0.5cm,minimum height=0.5cm},column sep={1cm,between origins},row sep={1cm,between origins}] (qmdd2) {
			\& |(m1)|$q_0$ 											\\
			\& |(m2a)| $q_1$                                 \&  						\\
			\& |(m3a)| $q_2$ 					\&  	\\
			\& |[terminal] (t3)| $1$ 									\\
		};
		
		\draw ($(m1)+(0,0.7cm)$) -- (m1) node[right, midway]{$\frac{1}{\sqrt{2}}$};
		
		\draw (m1) -- ++(240:0.6cm) -- (m2a);
		\draw (m1) -- ++(300:0.4cm) node[below, xshift=0.5pt, inner sep=0,font=\tiny] {$0$};
		
		\draw (m2a) -- ++(240:0.6cm) -- (m3a);
		\draw (m2a) -- ++(300:0.6cm) -- (m3a);
		
		\draw (m3a) -- ++(240:0.4cm) node[below, xshift=0.5pt, inner sep=0,font=\tiny] {$0$};
		\draw (m3a) -- ++(300:0.6cm) -- (t3);
		\end{tikzpicture}}
		\caption{Decision diagram}
		\label{fig:dd-statevector-approximated-dd}
	\end{subfigure}
	\caption{Approximated quantum state}
	\label{fig:dd-statevector-approximated}
\end{figure}
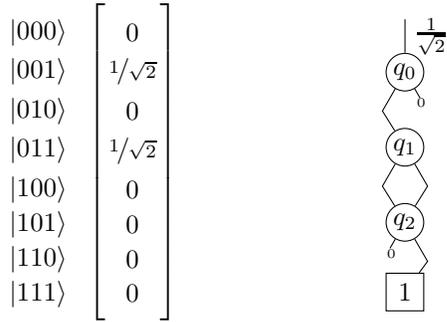

Approximating quantum states requires a metric to quantify how much the resulting state deviates from the original one---describing the effect of the approximation and whether the obtained results are useful.
Here, we use the \emph{fidelity metric}~\cite{NC:2000}, which describes the similarity between two quantum states. 
The fidelity of two pure quantum states \(\ket{\varphi}\) and \(\ket{\psi}\) is defined by
\begin{align}
	F(\ket{\varphi}, \ket{\psi}) = \left|\langle\varphi\ket{\psi}\right|^2 = \left| \left(\varphi^*\right)^{\smash{\mathrm{T}}} \cdot \psi \right|^2. \label{eq:fidelity}
\end{align}
The fidelity metric describes the likelihood that both quantum states produce equal outcomes after they are measured. Further, this metric is preserved when applying unitary operations.
The fidelity metric can be efficiently computed in terms of decision diagrams, i.e.,~in a linear fashion with respect to the number of DD~nodes.

\addtocounter{example}{-1}
\begin{example}[continued]
	Consider again the decision diagrams shown in Fig.~\ref{fig:dd-statevector} and Fig.~\ref{fig:dd-statevector-approximated-dd} (representing the original and the approximated quantum state \(\ket{\varphi}\) and \(\ket{\psi}\), respectively).
	The fidelity according to Eq.~(\ref{eq:fidelity}) of these two quantum states is computed as
	\begin{align*}F(\ket{\varphi}, \ket{\psi}) = \left|\left(\varphi^*\right)^{\smash{\mathrm{T}}} \cdot \psi\right|^2 = \left| \frac{2}{\sqrt{10}} \cdot \frac{1}{\sqrt{2}} + \frac{2}{\sqrt{10}} \cdot \frac{1}{\sqrt{2}}\right|^2 = \frac{4}{5}.\end{align*}
	That is, in \SI{80}{\percent} of the cases, a measurement of both states yield the same result. 
	Hence, the size of the decision diagram is reduced by \SI{50}{\percent}, while 
	the fidelity was only reduced by \SI{20}{\percent}.
\end{example}

The above observations illustrate the potential of approximating quantum states to obtain more compact \mbox{DD-based} representations 
and motivate to further explore this direction. 
To this end, the next section proposes several approximation schemes that explicitly trade off state fidelity for memory reduction.

\section{Approximation Schemes}
\label{sec:approx_schemes}

This section introduces four schemes to approximate a quantum state based on the ideas described in Section~\ref{sec:gen_idea}---two based on a repeated traversal and two based on considerations regarding the impact of the sub-vectors represented in the decision diagram.
These schemes aim to identify DD-nodes that do not significantly contribute to the overall state vector, eliminate them (i.e.,~setting certain small amplitudes to zero), and, by this, yield more compact representations.

\subsection{Approximation based on Traversal}

The general idea of \emph{approximation based on traversal} is to sample \(L\) bitstrings according to the probability distribution described by the state vector. A bitstring is sampled by following a path from the root node of the decision diagram to the terminal. Taking the left (right) successor of a node results in basis state \ket{0} (\ket{1}) for the respective qubit. To sample according to the probability distribution given by the quantum state, the \emph{upstream probabilities} are determined for each node. 
The upstream probability of a node (determining the summed probability of all paths from the node to the terminal) is computed in a depth-first fashion as sum of the upstream probabilities of the successor nodes---weighted by the squared magnitude
of the weight attached to the connecting edge. Then, the probability for choosing the right (left) successor is given by the upstream probability of these nodes (again weighted by the attached weight of the connecting edge). Since this is similar to measuring a quantum state (without collapsing it to a basis state), we refer to~\cite{DBLP:journals/tcad/ZulehnerW18} for further details.

Nodes that are not traversed during this sampling procedure hardly contribute to the overall quantum state, and are hence eliminated (i.e.,~replaced by zero-stubs) while keeping the fidelity of the approximated state close to one (assuming a sufficiently large number of samples \(L\) is chosen). This results in an approximation scheme that requires $\mathcal{O}(L\cdot n)$ time since each traversal visits the DD nodes from the root node to the terminal. Calculating the upstream probabilities requires to visit each node, and has to be done only once before starting to draw samples.

\begin{example}\label{ex:schemes-traversal}
Consider again Fig.~\ref{fig:dd-statevector}. The left (right) successor of node \(q_0\) has a probability of \num{0.8} (\num{0.2}) to be chosen during traversals.
Assuming the number of traversals \(L=3\), a possible distribution of traversed paths contains basis states \ket{001} (twice) and \ket{011} (once).
Therefore the nodes visited on these paths are kept, whereas the right-successor node \(q_1\) and in turn its successors labeled \(q_2\) are eliminated from the decision diagram.
The traversed nodes are sketched in Fig.~\ref{fig:traversals-visited} (the eliminated nodes are shown with a dashed outline). 
Eliminating the remaining node results in the approximated decision diagram shown in Fig.~\ref{fig:dd-statevector-approximated-dd} and a fidelity of \SI{80}{\percent} with respect to the original quantum state. 
\end{example}

\subsection{Approximation based on Traversal with Threshold}

The traversal scheme above can be generalized by counting how many times a node has been visited.
Additionally, we introduce a threshold \(\tau\) and only keep nodes that are visited more than \(\tau\) times, so that nodes visited  $\tau$ times or less are eliminated.
The idea of the threshold is to filter \enquote{outliers} that do not contribute significantly to the quantum state but are nonetheless encountered by chance during the traversals.

\begin{figure}[tbp]
	\begin{subfigure}[t]{0.3\linewidth}
		\resizebox{\linewidth}{!}{\begin{tikzpicture}[terminal/.style={draw,rectangle,inner sep=0pt}]	
		\matrix[matrix of nodes,ampersand replacement=\&,every node/.style={draw,circle,inner sep=0pt,minimum width=0.5cm,minimum height=0.5cm},column sep={1cm,between origins},row sep={1cm,between origins}] (qmdd2) {
							\& |[thick,fill=lightgray!25](m1)|$q_0$ 											\\
			|[thick,fill=lightgray!25](m2a)| $q_1$	\&                                  \& |[xshift=-0.5cm,dashed](m2b)| $q_2$ 						\\
			|[thick,fill=lightgray!25](m3a)| $q_2$	\& |[xshift=0.25cm,dashed](m3b)| $q_2$ 					\& |[dashed](m3c)| $q_2$ 	\\
							\& |[terminal] (t3)|  \(1\)									\\
		};
		
		\draw ($(m1)+(0,0.7cm)$) -- (m1);
		
		\draw (m1) -- ++(240:0.6cm) -- (m2a);
		\draw[dashed] (m1) -- ++(300:0.6cm) -- (m2b);
		
		\draw (m2a) -- ++(240:0.6cm) -- (m3a);
		\draw (m2a) -- ++(300:0.6cm) -- (m3a);
		
		\draw[dashed] (m2b) -- ++(240:0.6cm) -- (m3b);
		\draw[dashed] (m2b) -- ++(300:0.6cm) -- (m3c);
		
		\draw (m3a) -- ++(240:0.4cm) node[below, xshift=0.5pt, inner sep=0,font=\tiny] {$0$};
		\draw (m3a) -- ++(300:0.6cm) -- (t3);
		
		\draw[dashed] (m3b) -- ++(240:0.6cm) -- (t3);
		\draw[dashed] (m3b) -- ++(300:0.4cm) node[below, xshift=0.5pt, inner sep=0,font=\tiny] {$0$};
				
		\draw[dashed] (m3c) -- ++(240:0.4cm) node[below, xshift=0.5pt, inner sep=0,font=\tiny] {$0$};
		\draw[dashed] (m3c) -- ++(300:0.6cm) -- (t3);
		\end{tikzpicture}}
		\caption{Nodes visited during 3 traversals}
		\label{fig:traversals-visited}
	\end{subfigure}
	\hfill
	\begin{subfigure}[t]{0.3\linewidth}
		\resizebox{\linewidth}{!}{\begin{tikzpicture}[terminal/.style={draw,rectangle,inner sep=0pt}]	
		\matrix[matrix of nodes,ampersand replacement=\&,every node/.style={draw,circle,inner sep=0pt,minimum width=0.5cm,minimum height=0.5cm},column sep={1cm,between origins},row sep={1cm,between origins}] (qmdd2) {
							\& |(m1)|$10$ 											\\
			|(m2a)| $8$	\&                                  \& |[xshift=-0.5cm,dashed](m2b)| $2$ 						\\
			|(m3a)| $8$	\& |[xshift=0.25cm,dashed](m3b)| $0$ 					\& |[dashed](m3c)| $2$ 	\\
							\& |[terminal] (t3)| \(1\) 									\\
		};
		
		\draw ($(m1)+(0,0.7cm)$) -- (m1);
		
		\draw (m1) -- ++(240:0.6cm) -- (m2a);
		\draw[dashed] (m1) -- ++(300:0.6cm) -- (m2b);
		
		\draw (m2a) -- ++(240:0.6cm) -- (m3a);
		\draw (m2a) -- ++(300:0.6cm) -- (m3a);
		
		\draw[dashed] (m2b) -- ++(240:0.6cm) -- (m3b);
		\draw[dashed] (m2b) -- ++(300:0.6cm) -- (m3c);
		
		\draw (m3a) -- ++(240:0.4cm) node[below, xshift=0.5pt, inner sep=0,font=\tiny] {$0$};
		\draw (m3a) -- ++(300:0.6cm) -- (t3);
		
		\draw[dashed] (m3b) -- ++(240:0.6cm) -- (t3);
		\draw[dashed] (m3b) -- ++(300:0.4cm) node[below, xshift=0.5pt, inner sep=0,font=\tiny] {$0$};
				
		\draw[dashed] (m3c) -- ++(240:0.4cm) node[below, xshift=0.5pt, inner sep=0,font=\tiny] {$0$};
		\draw[dashed] (m3c) -- ++(300:0.6cm) -- (t3);
		\end{tikzpicture}}
		\caption{Number of times nodes are visited for 10 traversals}
		\label{fig:traversals-count}
	\end{subfigure}
	\hfill
	\begin{subfigure}[t]{0.33\linewidth}
		\centering
		\resizebox{0.90\linewidth}{!}{\begin{tikzpicture}[terminal/.style={draw,rectangle,inner sep=0pt}]	
			\matrix[matrix of nodes,ampersand replacement=\&,every node/.style={draw,circle,inner sep=0pt,minimum width=0.5cm,minimum height=0.5cm},column sep={1cm,between origins},row sep={1cm,between origins}] (qmdd2) {
				\& |(m1)|\(1.0\) \\
				|(m2a)| $0.8$ \& \& |[xshift=-0.5cm,dashed](m2b)| $0.2$ \\
				|(m3a)| $0.8$ \& |[xshift=0.25cm,dashed](m3b)| $0.1$ \& |[dashed](m3c)| $0.1$ \\
				\& |[terminal] (t3)| $1$ \\
			};
			
			\draw ($(m1)+(0,0.7cm)$) -- (m1);
			
			\draw (m1) -- ++(240:0.6cm) -- (m2a);
			\draw[dashed] (m1) -- ++(300:0.6cm) -- (m2b);
			
			\draw (m2a) -- ++(240:0.6cm) -- (m3a);
			\draw (m2a) -- ++(300:0.6cm) -- (m3a);
			
			\draw[dashed] (m2b) -- ++(240:0.6cm) -- (m3b);
			\draw[dashed] (m2b) -- ++(300:0.6cm) -- (m3c);
			
			\draw (m3a) -- ++(240:0.4cm) node[below, xshift=0.5pt, inner sep=0,font=\tiny] {$0$};
			\draw (m3a) -- ++(300:0.6cm) -- (t3);
			
			\draw[dashed] (m3b) -- ++(240:0.6cm) -- (t3);
			\draw[dashed] (m3b) -- ++(300:0.4cm) node[below, xshift=0.5pt, inner sep=0,font=\tiny] {$0$};
					
			\draw[dashed] (m3c) -- ++(240:0.4cm) node[below, xshift=0.5pt, inner sep=0,font=\tiny] {$0$};
			\draw[dashed] (m3c) -- ++(300:0.6cm) -- (t3);
		\end{tikzpicture}}
		\caption{Proportional contribution to the state for each node per level}
		\label{fig:fixedfidelity-downstream}
	\end{subfigure}
	\caption{Illustration of approximation schemes with nodes to be eliminated shown with a dashed outline}
\end{figure}
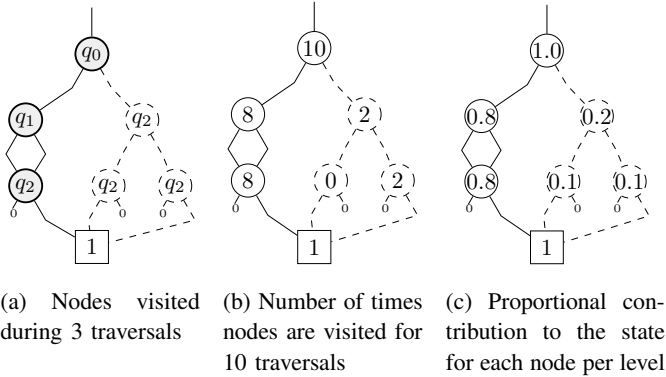

\begin{example}
Consider again Fig.~\ref{fig:dd-statevector} and Example~\ref{ex:schemes-traversal}.
Increasing the number of traversals to \(L=10\) may lead to the possible distribution of following the left (right) successor of \(q_0\) 8~times (2~times) as illustrated in Fig.~\ref{fig:traversals-count}.
This distribution is closer to the actual probabilities but also includes the right-hand nodes \(q_1\) and \(q_2\).
Introducing a threshold on how many times a node has to be visited during the traversals serves as a cut-off point for nodes with a lower probabilities. Selecting a threshold of \(\tau=3\) leads to the elimination of the dashed nodes and, hence, to the approximation depicted in Fig.~\ref{fig:dd-statevector-approximated-dd}.
\end{example}

\subsection{Approximation with a Target Fidelity}
\label{subsec:fixed-fidelity}

As discussed in the previous section, it is desirable to approximate quantum states with the fidelity desired in a given application context. The traversal-based approximation schemes so far do not provide this feature. Hence, we present further schemes that guarantee the minimum fidelity of the approximated state.

To this end, the entire decision diagram is traversed twice: once in a depth-first fashion to compute the \emph{upstream probabilities} (as discussed above) and once in a breath-first fashion to compute the \emph{downstream probabilities} (i.e.,~the summed probabilities of all paths from the root node to the considered one). The downstream probability of a node is recursively computed as a sum of the downstream probabilities of its parents---weighted by the squared magnitude of the weight attached to the connecting edge.

To guarantee a target fidelity $f$, a level is chosen and the nodes of this level are sorted by their contribution for the overall state vector in ascending order.
The contribution is calculated by multiplying the downstream probabilities of the nodes with the summed upstream probabilities of the successors (considering again the weight attached to the connecting edge). 
These numbers sum up to 1 since a quantum state has a norm of one. An approximated state vector with fidelity $f$ is now determined by summing up the sorted significance of the nodes until the sum exceeds $1-f$, 
followed by the elimination of all nodes included in the sum so far. Eliminating only these nodes guarantees that the approximated state vector has at least fidelity $f$.

The difficulty in this approach is to choose the level from which the nodes shall be eliminated in order to obtain the most compact decision diagram. 
One possibility to tackle this issue is to precompute how many nodes can be eliminated at each level. However, this still does not allow predicting how the elimination of nodes affects nodes below the chosen level. In fact, computing how the elimination of certain nodes affects the DD size requires determining all nodes to be removed.

\begin{example}
Consider again Fig.~\ref{fig:dd-statevector}.
The calculated contributions based on upstream and downstream probabilities are depicted in Fig.~\ref{fig:fixedfidelity-downstream}.
Given a target fidelity of \num{0.5} and choosing the second level (i.e., the nodes labeled \(q_1\)), the probabilities in ascending order are \num{0.2} and \num{0.8}.
The sum \(0.2 + 0.8\) exceeds \(0.5\), therefore only the \(q_1\) node on the right will be eliminated including its both successors (shown with a dashed outline)---leading to the approximation depicted in Fig.~\ref{fig:dd-statevector-approximated-dd}.
\end{example}

\subsection{Approximation with a Target Fidelity per Level}
\label{subsec:fixed-levels}

Approximating a quantum state by targeting the fidelity of a single level requires determining a suitable level to eliminate nodes.
As mentioned earlier, a possible solution is to calculate the fidelities for each level and choose one.
Assuming that the target fidelity \(f\) is not required to be the absolute minimum fidelity acceptable, the visited nodes of each level can be eliminated instead of only eliminating the nodes for one level.
This cannot guarantee fidelity $f$, but the resulting fidelity admits a lower bound $f^{n-1}$ with \(n\) denoting the number of qubits. Notably, this bound was hardly reached in our experiments.

\begin{example}
Consider again Fig.~\ref{fig:dd-statevector} and the associated contributions to the quantum state depicted in Fig.~\ref{fig:fixedfidelity-downstream}.
To apply the fidelity-per-level scheme, the downstream probabilities are summed up for each level until reaching the target of \(0.5\) and all nodes visited up to this point are eliminated.
More precisely, for the first level \(q_0\) the single node cannot be eliminated as it contributes to every basis state. 
On the second level the right node \(q_1\) is eliminated as in the previous example.
Finally, on the third level the two nodes \(q_2\) on the right are eliminated as they only contribute \SI{20}{\percent} to this level.\footnote{This step is kept in the example as these nodes would have been eliminated automatically along with the \(q_1\) node.}
These eliminations (i.e,~the nodes drawn with a dashed outline) result in the approximation depicted in Fig.~\ref{fig:dd-statevector-approximated-dd}.
\end{example}

\begin{figure*}[tbp]
	\pgfplotsset{
		every axis legend/.append style={draw=none, fill=none, font=\small},
		legend entry/.initial=,
		every axis plot post/.code={%
			\pgfkeysgetvalue{/pgfplots/legend entry}\tempValue
			\ifx\tempValue\empty
			\pgfkeysalso{/pgfplots/forget plot}%
			\else
			\expandafter\addlegendentry\expandafter{\tempValue}%
			\fi
		},
	}
	\begin{minipage}[t]{0.49\linewidth}
		\centering
		\begin{subfigure}[t]{0.48\linewidth}
			\resizebox{\linewidth}{!}{\begin{tikzpicture}
				\begin{axis}[
				xlabel=Traversals,
				ylabel=Attained fidelity, 
				ylabel near ticks, xlabel near ticks,
				legend style={at={(0.95,0.05)},anchor=south east},
				ymin=0, ymax=1.1, ytickmax=1,
				]
				\addplot[teal, legend entry={supremacy\_4x4\_10}] table[col sep=comma, x=samples, y=fidelity] {./pgfplot/sampling-inst-4x4-10.csv};
				\addplot[teal, legend entry={supremacy\_4x4\_15}] table[col sep=comma, x=samples, y=fidelity] {./pgfplot/sampling-inst-4x4-15.csv};
				\addplot[teal, legend entry={supremacy\_5x4\_10}] table[col sep=comma, x=samples, y=fidelity] {./pgfplot/sampling-inst-5x4-10.csv};
				\addplot[orange, legend entry={shor\_33\_2}] table[col sep=comma, x=samples, y=fidelity] {./pgfplot/sampling-shor-33-2.csv};
				\addplot[orange, legend entry={shor\_55\_2}] table[col sep=comma, x=samples, y=fidelity] {./pgfplot/sampling-shor-55-2.csv};
				\addplot[orange, legend entry={shor\_69\_4}] table[col sep=comma, x=samples, y=fidelity] {./pgfplot/sampling-shor-69-4.csv};
				\addplot[orange, legend entry={shor\_221\_4}] table[col sep=comma, x=samples, y=fidelity] {./pgfplot/sampling-shor-221-4.csv};
				\addplot[blue, legend entry={qua\_chem\_3x3}] table[col sep=comma, x=samples, y=fidelity] {./pgfplot/sampling-jellium-3x3.csv};
				\end{axis}
				\end{tikzpicture}}
			\newlength{\somelength}\settowidth{\somelength}{(a)~}%
			\caption{\resizebox{\linewidth-\somelength}{!}{Attained\,fidelity\,v.\,\#traversals}}
			\label{fig:approx_by_sampling-fidelity}
		\end{subfigure}
		\hfill
		\begin{subfigure}[t]{0.48\linewidth}
			\resizebox{\linewidth}{!}{\begin{tikzpicture}
				\begin{axis}[
				xlabel=Traversals,
				ylabel=Relative DD size, 
				ylabel near ticks, xlabel near ticks,
				legend style={at={(0.05,0.95)},anchor=north west},
				ymin=0, ymax=1.1, ytickmax=1,
				]
				\addplot[teal, legend entry={supremacy\_4x4\_10}] table[col sep=comma, x=samples, y=relddsize] {./pgfplot/sampling-inst-4x4-10.csv};
				\addplot[teal, legend entry={supremacy\_4x4\_15}] table[col sep=comma, x=samples, y=relddsize] {./pgfplot/sampling-inst-4x4-15.csv};
				\addplot[teal, legend entry={supremacy\_5x4\_10}] table[col sep=comma, x=samples, y=relddsize] {./pgfplot/sampling-inst-5x4-10.csv};
				\addplot[orange, legend entry={shor\_33\_2}] table[col sep=comma, x=samples, y=relddsize] {./pgfplot/sampling-shor-33-2.csv};
				\addplot[orange, legend entry={shor\_55\_2}] table[col sep=comma, x=samples, y=relddsize] {./pgfplot/sampling-shor-55-2.csv};
				\addplot[orange, legend entry={shor\_69\_4}] table[col sep=comma, x=samples, y=relddsize] {./pgfplot/sampling-shor-69-4.csv};
				\addplot[orange, legend entry={shor\_221\_4}] table[col sep=comma, x=samples, y=relddsize] {./pgfplot/sampling-shor-221-4.csv};
				\addplot[blue, legend entry={qua\_chem\_3x3}] table[col sep=comma, x=samples, y=relddsize] {./pgfplot/sampling-jellium-3x3.csv};
				\end{axis}
				\end{tikzpicture}}
			\caption{Size v. \#traversals}
			\label{fig:approx_by_sampling-size}
		\end{subfigure}
		\caption{Approximation based on the number of traversals}
		\label{fig:approx_by_sampling}
		\vspace{0.5em}
	\end{minipage}
	\hfill
	\begin{minipage}[t]{0.49\linewidth}
		\centering
		\begin{subfigure}[t]{0.48\linewidth}
			\resizebox{\linewidth}{!}{\begin{tikzpicture}
				\begin{axis}[
				xlabel=Threshold,
				ylabel=Attained fidelity, 
				ylabel near ticks, xlabel near ticks,
				legend style={at={(0.5,0.05)},anchor=south},
				ymin=0, ymax=1.1, ytickmax=1,
				]
				\addplot[teal, legend entry={supremacy\_4x4\_10}] table[col sep=comma, x=threshold, y=relddsize] {./pgfplot/threshold-inst-4x4-10.csv};
				\addplot[teal, legend entry={supremacy\_4x4\_15}] table[col sep=comma, x=threshold, y=relddsize] {./pgfplot/threshold-inst-4x4-15.csv};
				\addplot[teal, legend entry={supremacy\_5x4\_10}] table[col sep=comma, x=threshold, y=relddsize] {./pgfplot/threshold-inst-5x4-10.csv};
				\addplot[orange, legend entry={shor\_33\_2}] table[col sep=comma, x=threshold, y=fidelity] {./pgfplot/threshold-shor-33-2.csv};
				\addplot[orange, legend entry={shor\_55\_2}] table[col sep=comma, x=threshold, y=fidelity] {./pgfplot/threshold-shor-55-2.csv};
				\addplot[orange, legend entry={shor\_69\_4}] table[col sep=comma, x=threshold, y=fidelity] {./pgfplot/threshold-shor-69-4.csv};
				\addplot[orange, legend entry={shor\_221\_4}] table[col sep=comma, x=threshold, y=fidelity] {./pgfplot/threshold-shor-221-4.csv};
				\addplot[blue, legend entry={qua\_chem\_3x3}] table[col sep=comma, x=threshold, y=fidelity] {./pgfplot/threshold-jellium-3x3.csv};
				\end{axis}
				\end{tikzpicture}}
			\caption{Attained fidelity v.\,threshold}
			\label{fig:approx_by_threshold_sampling-fidelity}
		\end{subfigure}
		\hfill
		\begin{subfigure}[t]{0.48\linewidth}
			\resizebox{\linewidth}{!}{\begin{tikzpicture}
				\begin{axis}[
				xlabel=Threshold,
				ylabel=Relative DD size, 
				ylabel near ticks, xlabel near ticks,
				legend style={at={(0.95,0.95)},anchor=north east},
				ymin=0, ymax=1.1, ytickmax=1,
				]
				\addplot[teal, legend entry={supremacy\_4x4\_10}] table[col sep=comma, x=threshold, y=relddsize] {./pgfplot/threshold-inst-4x4-10.csv};
				\addplot[teal, legend entry={supremacy\_4x4\_15}] table[col sep=comma, x=threshold, y=relddsize] {./pgfplot/threshold-inst-4x4-15.csv};
				\addplot[teal, legend entry={supremacy\_5x4\_10}] table[col sep=comma, x=threshold, y=relddsize] {./pgfplot/threshold-inst-5x4-10.csv};
				\addplot[orange, legend entry={shor\_33\_2}] table[col sep=comma, x=threshold, y=relddsize] {./pgfplot/threshold-shor-33-2.csv};
				\addplot[orange, legend entry={shor\_55\_2}] table[col sep=comma, x=threshold, y=relddsize] {./pgfplot/threshold-shor-55-2.csv};
				\addplot[orange, legend entry={shor\_69\_4}] table[col sep=comma, x=threshold, y=relddsize] {./pgfplot/threshold-shor-69-4.csv};
				\addplot[orange, legend entry={shor\_221\_4}] table[col sep=comma, x=threshold, y=relddsize] {./pgfplot/threshold-shor-221-4.csv};
				\addplot[blue, legend entry={qua\_chem\_3x3}] table[col sep=comma, x=threshold, y=relddsize] {./pgfplot/threshold-jellium-3x3.csv};
				\end{axis}
				\end{tikzpicture}}
			\caption{Size v. threshold}
			\label{fig:approx_by_threshold_sampling-size}
		\end{subfigure}
		\caption{Approximation based on traversals with a threshold}
		\label{fig:approx_by_threshold_sampling}
	\end{minipage}
	
	\vspace{\floatsep}
	\begin{minipage}[t]{0.49\linewidth}
		\centering
		\begin{subfigure}[t]{0.48\linewidth}
			\resizebox{\linewidth}{!}{\begin{tikzpicture}
				\begin{axis}[
				xlabel=Target fidelity,
				ylabel=Attained fidelity, 
				ylabel near ticks, xlabel near ticks,
				legend style={at={(0.05,0.95)},anchor=north west},
				ymin=0, ymax=1.1, ytickmax=1,
				xmin=0, xmax=1.1, xtickmax=1,
				]
				\addplot[teal, legend entry={supremacy\_4x4\_10}] table[col sep=comma, x=target, y=fidelity] {./pgfplot/fixedfidelity-inst-4x4-10.csv};
				\addplot[teal, legend entry={supremacy\_4x4\_15}] table[col sep=comma, x=target, y=fidelity] {./pgfplot/fixedfidelity-inst-4x4-15.csv};
				\addplot[teal, legend entry={supremacy\_5x4\_10}] table[col sep=comma, x=target, y=fidelity] {./pgfplot/fixedfidelity-inst-5x4-10.csv};
				\addplot[orange, legend entry={shor\_33\_2}] table[col sep=comma, x=target, y=fidelity] {./pgfplot/fixedfidelity-shor-33-2.csv};
				\addplot[orange, legend entry={shor\_55\_2}] table[col sep=comma, x=target, y=fidelity] {./pgfplot/fixedfidelity-shor-55-2.csv};
				\addplot[orange, legend entry={shor\_69\_4}] table[col sep=comma, x=target, y=fidelity] {./pgfplot/fixedfidelity-shor-69-4.csv};
				\addplot[orange, legend entry={shor\_221\_4}] table[col sep=comma, x=target, y=fidelity] {./pgfplot/fixedfidelity-shor-221-4.csv};
				\addplot[blue, legend entry={qua\_chem\_3x3}] table[col sep=comma, x=target, y=fidelity] {./pgfplot/fixedfidelity-jellium-3x3.csv};
				\end{axis}
				\end{tikzpicture}}
			\caption{Attained fidelity v. target}
			\label{fig:approx_fixed_fidelity-fidelity}
		\end{subfigure}
		\hfill
		\begin{subfigure}[t]{0.48\linewidth}
			\resizebox{\linewidth}{!}{\begin{tikzpicture}
				\begin{axis}[
				xlabel=Target fidelity,
				ylabel=Relative DD size, 
				ylabel near ticks, xlabel near ticks,
				legend style={at={(0.05,0.95)},anchor=north west},
				ymin=0, ymax=1.1, ytickmax=1,
				xmin=0, xmax=1.1, xtickmax=1,
				]
				\addplot[teal, legend entry={supremacy\_4x4\_10}] table[col sep=comma, x=target, y=relddsize] {./pgfplot/fixedfidelity-inst-4x4-10.csv};
				\addplot[teal, legend entry={supremacy\_4x4\_15}] table[col sep=comma, x=target, y=relddsize] {./pgfplot/fixedfidelity-inst-4x4-15.csv};
				\addplot[teal, legend entry={supremacy\_5x4\_10}] table[col sep=comma, x=target, y=relddsize] {./pgfplot/fixedfidelity-inst-5x4-10.csv};
				\addplot[orange, legend entry={shor\_33\_2}] table[col sep=comma, x=target, y=relddsize] {./pgfplot/fixedfidelity-shor-33-2.csv};
				\addplot[orange, legend entry={shor\_55\_2}] table[col sep=comma, x=target, y=relddsize] {./pgfplot/fixedfidelity-shor-55-2.csv};
				\addplot[orange, legend entry={shor\_69\_4}] table[col sep=comma, x=target, y=relddsize] {./pgfplot/fixedfidelity-shor-69-4.csv};
				\addplot[orange, legend entry={shor\_221\_4}] table[col sep=comma, x=target, y=relddsize] {./pgfplot/fixedfidelity-shor-221-4.csv};
				\addplot[blue, legend entry={qua\_chem\_3x3}] table[col sep=comma, x=target, y=relddsize] {./pgfplot/fixedfidelity-jellium-3x3.csv};
				\end{axis}
				\end{tikzpicture}}
			\caption{Size v. target fidelity}
			\label{fig:approx_fixed_fidelity-size}
		\end{subfigure}
		\caption{Approximation for a target fidelity}
		\label{fig:approx_fixed_fidelity}
	\end{minipage}
	\hfill
	\begin{minipage}[t]{0.49\linewidth}
		\centering
		\begin{subfigure}[t]{0.48\linewidth}
			\resizebox{\linewidth}{!}{\begin{tikzpicture}
				\begin{axis}[
				xlabel=Target fidelity,
				ylabel=Attained fidelity, 
				ylabel near ticks, xlabel near ticks,
				legend style={at={(0.05,0.95)},anchor=north west},
				ymin=0, ymax=1.1, ytickmax=1,
				xmin=0, xmax=1.1, xtickmax=1,
				]
				\addplot[teal, legend entry={supremacy\_4x4\_10}] table[col sep=comma, x=target, y=fidelity] {./pgfplot/fixedlevels-inst-4x4-10.csv};
				\addplot[teal, legend entry={supremacy\_4x4\_15}] table[col sep=comma, x=target, y=fidelity] {./pgfplot/fixedlevels-inst-4x4-15.csv};
				\addplot[teal, legend entry={supremacy\_5x4\_10}] table[col sep=comma, x=target, y=fidelity] {./pgfplot/fixedlevels-inst-5x4-10.csv};
				\addplot[orange, legend entry={shor\_33\_2}] table[col sep=comma, x=target, y=fidelity] {./pgfplot/fixedlevels-shor-33-2.csv};
				\addplot[orange, legend entry={shor\_55\_2}] table[col sep=comma, x=target, y=fidelity] {./pgfplot/fixedlevels-shor-55-2.csv};
				\addplot[orange, legend entry={shor\_69\_4}] table[col sep=comma, x=target, y=fidelity] {./pgfplot/fixedlevels-shor-69-4.csv};
				\addplot[orange, legend entry={shor\_221\_4}] table[col sep=comma, x=target, y=fidelity] {./pgfplot/fixedlevels-shor-221-4.csv};
				\addplot[blue, legend entry={qua\_chem\_3x3}] table[col sep=comma, x=target, y=fidelity] {./pgfplot/fixedlevels-jellium-3x3.csv};
				\end{axis}
				\end{tikzpicture}}
			\caption{Attained fidelity v. target}
			\label{fig:approx_level_fidelity-fidelity}
		\end{subfigure}
		\hfill
		\begin{subfigure}[t]{0.48\linewidth}
			\resizebox{\linewidth}{!}{\begin{tikzpicture}
				\begin{axis}[
				xlabel=Target fidelity,
				ylabel=Relative DD size, 
				ylabel near ticks, xlabel near ticks,
				legend style={at={(0.05,0.95)},anchor=north west},
				ymin=0, ymax=1.1, ytickmax=1,
				xmin=0, xmax=1.1, xtickmax=1,
				]
				\addplot[teal, legend entry={supremacy\_4x4\_10}] table[col sep=comma, x=target, y=relddsize] {./pgfplot/fixedlevels-inst-4x4-10.csv};
				\addplot[teal, legend entry={supremacy\_4x4\_15}] table[col sep=comma, x=target, y=relddsize] {./pgfplot/fixedlevels-inst-4x4-15.csv};
				\addplot[teal, legend entry={supremacy\_5x4\_10}] table[col sep=comma, x=target, y=relddsize] {./pgfplot/fixedlevels-inst-5x4-10.csv};
				\addplot[orange, legend entry={shor\_33\_2}] table[col sep=comma, x=target, y=relddsize] {./pgfplot/fixedlevels-shor-33-2.csv};
				\addplot[orange, legend entry={shor\_55\_2}] table[col sep=comma, x=target, y=relddsize] {./pgfplot/fixedlevels-shor-55-2.csv};
				\addplot[orange, legend entry={shor\_69\_4}] table[col sep=comma, x=target, y=relddsize] {./pgfplot/fixedlevels-shor-69-4.csv};
				\addplot[orange, legend entry={shor\_221\_4}] table[col sep=comma, x=target, y=relddsize] {./pgfplot/fixedlevels-shor-221-4.csv};
				\addplot[blue, legend entry={qua\_chem\_3x3}] table[col sep=comma, x=target, y=relddsize] {./pgfplot/fixedlevels-jellium-3x3.csv};
				\end{axis}
				\end{tikzpicture}}
			\caption{Size v. target fidelity}
			\label{fig:approx_level_fidelity-size}
		\end{subfigure}
		\caption{Approximation for a target fidelity per level}
		\label{fig:approx_level_fidelity}
		\vspace{0.5em}
	\end{minipage}
\end{figure*}

\section{Evaluation}
\label{sec:evaluation}

In order to evaluate the potential of approximation for obtaining more compact DD-based representations of quantum states, we implemented the four schemes discussed in Section~\ref{sec:approx_schemes} (on top of \cite{zulehner2019howtoefficiently}). Afterwards, the respectively approximated decision diagrams have been evaluated regarding their resulting fidelity and the size compared to the original (i.e.,~\mbox{non-approximated}) decision diagrams. 
To this end, quantum states generated by representative quantum functionalities from the following groups have been considered as benchmarks: 
\begin{itemize}
	\item Quantum circuits provided by researchers from Google~\cite{boixo2016characterizing} as candidates to establish quantum-computational supremacy using \mbox{controlled-phase} gates (denoted \enquote{supremacy\_\(A\)x\(B\)\_\(C\)}, representing a circuit on an \(A\times B\) surface with depth \(C\)),
	\item quantum circuits realizing Shor's algorithm to factorize integers~\cite{Sho:94} (denoted \enquote{shor\_\(A\)\_\(B\)} for factorizing \(A\) with the coprime value \(B\)), and
	\item a quantum circuit simulating the uniform electron gas (taken from~\cite{PhysRevX.8.011044} and denoted \enquote{qua\_chem\_3x3}).
\end{itemize}

In the following, we present our main empirical results and put them in perspective.

\subsection{Comparison of Approximation Schemes}
\label{subsec:comparison}

In a first series of evaluations, the different approximation schemes are compared against each other. To this end, Figures~\ref{fig:approx_by_sampling} to~\ref{fig:approx_level_fidelity} show 
the respectively \emph{attained fidelity} and the \emph{relative DD size} of the approximated decision diagrams (compared to the original decision diagrams) for each of the four approximation schemes.
Here, the plots nicely show the overall trade-off between the attained fidelity and the DD size as well as the effect of the respective parameter associated with each scheme.
More precisely,
Fig.~\ref{fig:approx_by_sampling} shows how the number of traversals relates to the attained fidelity and relative DD size (more traversals yield more accurate decision diagrams which increase the fidelity but also the resulting DD size).
Fig.~\ref{fig:approx_by_threshold_sampling} demonstrates the same effect with a decreasing threshold, i.e.,~the smaller the threshold the more accurate (but also the larger) the decision diagrams.\footnote{One million traversals have been carried out before the threshold has been applied.}
Fig.~\ref{fig:approx_fixed_fidelity} shows that the scheme in Section~\ref{subsec:fixed-fidelity} always guaranteed the target fidelity (besides that, the smaller the target fidelity is given the smaller decision diagrams are obtained). 
Finally, Fig.~\ref{fig:approx_level_fidelity} shows that the scheme in Section~\ref{subsec:fixed-levels} might not always guarantee the target fidelity but still yields fidelities which are close to the targeted one.\footnote{Fig.~\ref{fig:approx_fixed_fidelity} and Fig.~\ref{fig:approx_level_fidelity} confirm that the corresponding schemes indeed approximate a quantum state towards a desired fidelity.} 

Overall, these results show that for \emph{all} schemes the DD size is greatly reduced with moderate effects on the fidelity in many cases. The actual effects significantly depend on the respectively considered quantum functionality. Quantum states obtained from~\emph{quantum-supremacy} circuits are the hardest to approximate (but some schemes at least lead to improvements which are proportional to the fidelity losses). This can be explained by the fact that \emph{most} amplitudes have comparable magnitudes in these cases and, hence, approximations harm the fidelity in the same proportion as the size is reduced. In contrast, quantum states arising in the remaining benchmarks frequently have many amplitudes which, probabilistically, are negligible and, hence, can easily be approximated. Eventually, this reduces the DD sizes to a fraction of its original size while controlling the attained fidelity for almost all schemes. This validates the initial idea proposed in this paper that approximating quantum states using decision diagrams is an effective technique for more compact representations.

\subsection{Application from a Designer's Perspective}

In a second series of evaluations, we additionally evaluated how the idea and findings proposed in this work explicitly aid quantum circuit designers and/or the development of corresponding design automation tools for quantum computing. 
Due to space limitations, we consider a representative scenario in which the acceptable fidelity is specified by the application context.\footnote{Assumptions about the \enquote{acceptable fidelity} of a result obtained from a quantum computer can be made based on the considered functionality~\cite{DBLP:journals/corr/abs-1807-10749}.} 
More precisely, assume that a fidelity of~0.5 is acceptable. 
Then, the designer can easily decide when to apply the scheme from Section~\ref{subsec:fixed-fidelity} to obtain an approximated quantum state, e.g.,~each time the decision diagram grows to big during simulation, in a fashion similar to garbage collection.
Further, since the fidelity metric of a sufficiently entangled quantum state is multiplicative, the overall fidelity after each step can be easily tracked.
Table~\ref{tbl:designer-eval} shows the corresponding characteristics of those decision diagrams (here, the name of the benchmark, the original DD size, the DD size of the approximated decision diagram, the relative size, and the fidelity of the approximated decision diagram is given).
These results show that exact quantum state representations can be replaced with approximations in practice (e.g.,~for purposes of simulation, verification, synthesis)---leading to reductions of the \mbox{DD size} of several orders of magnitude while, at the same time, controlling the attained fidelity. In the best case, this reduces the DD size from over a million nodes to just under 200 nodes.

\begin{table}
	\renewcommand{\tabcolsep}{3pt}
	\caption{Target fidelity scheme targeting \num{0.5} fidelity}
	\label{tbl:designer-eval}
	\scriptsize
	\centering
	\begin{tabular}{lS[table-format=7.0]S[table-format=6.0]*{2}{S[table-auto-round, table-format=1.4]}}
		\toprule
		Benchmark          & {Orig.~size} & {Approx.~size} & {Compression} & {Fidelity} \\ \midrule
		supremacy\_4x4\_10 & 65071      & 23305      & 0.358147    & 0.500019   \\
		supremacy\_4x4\_15 & 65536      & 24592      & 0.375244    & 0.50001    \\
		supremacy\_5x4\_10 & 482408     & 127412     & 0.264117    & 0.500021   \\[3pt]
		shor\_33\_2        & 48980      & 2115       & 0.0431809   & 0.524921   \\
		shor\_55\_2        & 93541      & 344        & 0.00367753  & 0.506433   \\
		shor\_69\_4        & 196234     & 16462      & 0.0838896   & 0.511702   \\
		shor\_221\_4       & 1048575    & 198        & 0.000188828 & 0.519653   \\[3pt]
		qua\_chem\_3x3     & 59594      & 147        & 0.00246669  & 0.509793   \\ \bottomrule
	\end{tabular}
\end{table}

\section{Conclusion}
\label{sec:conclusion}

In this work, we proposed and evaluated the idea of approximating quantum states using decision diagrams. 
To this end, we exploited the inherent error resistance in quantum algorithms, i.e.,~small inaccuracies of the state representation do not (or hardly) affect the measurement outcome, in order to approximate the state of a quantum system.
We applied this idea to decision diagrams by carefully analyzing their structure to quantify the contribution of each node to the quantum state.
This resulted in four dedicated approximation schemes---leading to a reduction of several orders of magnitude in DD size while, at the same time, controlling the obtained fidelity. 
Moreover, our best approximation schemes closely match the desired quantum fidelity or even guarantee it.
The findings of this work provide the basis for several design automation tools, e.g.,~for simulation, for verification, and for synthesis, that previously suffered from unaffordable memory requirements and may become practical at greater scales through the use of approximate quantum state representations in a fashion similar to garbage collection.

{
\section*{Acknowledgments}
This work has partially been supported by the LIT Secure and Correct Systems Lab funded by the State of Upper Austria.

\bibliographystyle{IEEEtran}
\let\oldbibliography\thebibliography
\renewcommand{\thebibliography}[1]{%
  \oldbibliography{#1}%
  \setlength{\itemsep}{1pt plus 3pt}%
}
\bibliography{../../bib/lit_header,../../bib/lit_misc,../../bib/lit_mymisc,../../bib/lit_myrev,../../bib/lit_others,../../bib/lit_othersrev,../../bib/lit_rev,../../bib/lit_new,../../bib/lit_adiabatic,../../bib/lit_memristor,../../bib/lit_simulation,../../bib/lit_quantum,./new-references} 
}
	
\end{document}